# First-principles study of the structural stability of Mn$_3$Z (Z=Ga, Sn and Ge) Heusler compounds

**Delin Zhang [1], Binghai Yan [1,2,3], Shu-Chun Wu [1], Jürgen Kübler [4], Guido Kreiner[1], Stuart S. P. Parkin [4], Claudia Felser [1,2]**

[1] Max-Planck-Institute für Chemische Physik fester Stoffe, Nöthnitzer Str. 40, 01187 Dresden, Germany

[2] Johannes Gutenberg-Universität Mainz, Staudingerweg 9, 55128 Mainz, Germany

[3] Max Planck Institute for the Physics of Complex Systems, Nöthnitzer Straße 38

01187 Dresden, Germany

[4] Institut für Festkörperphysik, Technische Universität Darmstadt, 64289 Darmstadt, Germany

[5] IBM Almaden Research Center San Jose, CA 95120, USA.

**Abstract**

We investigate the structural stability and magnetic properties of cubic, tetragonal and hexagonal phases of Mn$_3$Z (Z=Ga, Sn and Ge) Heusler compounds using first-principles density-functional theory. We propose that the cubic phase plays an important role as an intermediate state in the phase transition from the hexagonal to the tetragonal phases. Consequently, Mn$_3$Ga and Mn$_3$Ge behave differently from Mn$_3$Sn, because the relative energies of the cubic and hexagonal phases are different. This result agrees with experimental observations from these three compounds. The weak ferromagnetism of the hexagonal phase and the perpendicular magnetocrystalline anisotropy of the tetragonal phase obtained in our calculations are also consistent with experiment.

# I. Introduction

The Mn$_3$Z (Z=Ga, Sn and Ge) type of Heusler compounds can have three different structural phases, where each phase exhibits different magnetic properties. The hexagonal phase has been known for decades. Mn atoms form a Kagome lattice in a plane with a Z atom in the center of the hexagon. Here, Mn atoms have a triangular antiferromagnetic (AFM) coupling with a weak net magnetic moment [1~8]. The cubic phase is the standard full Heusler structure. In this phase, Mn atoms are present in two unique lattice sites; these different sites have magnetic moments with opposite directions, leading to ferrimagnetic (FiM) order [9, 10]. The cubic phase has a high density of states at the Fermi energy, and hence a Peierls transition could occur [11], giving rise to the third phase, the tetragonal phase. The tetragonal phase can be treated as a cubic phase with a distortion along the $z$ direction. This distortion makes the magnetic moments favor the $z$ axis, meaning the system possesses perpendicular magnetocrystalline anisotropy (PMA) [11], which promises great potential for future high-density spin-transfer torque applications [12,13,14~21].

In experiments, the hexagonal phase of Mn$_3$Z (Z=Ga, Sn and Ge) was synthesized by annealing the samples at high temperatures [1~3, 22], while the tetragonal phase of Mn$_3$Ga and Mn$_3$Ge was realized by low temperature annealing [1, 3, 23, 24]. The cubic phase has not been observed so far, since it may be unstable as mentioned above. On the other hand, the tetragonal phase of Mn$_3$Sn has not yet been reported in the literature, although it is expected to behave similarly to the other two compounds. The possible transition between hexagonal, cubic and tetragonal phases catches the attention of researchers [3, 25], but a comprehensive study is yet to be undertaken.

In this paper, we investigate the structural stability and magnetic properties of the hexagonal, cubic and tetragonal phases for the three compounds Mn$_3$Z (Z=Ga, Sn and Ge) using first-principles calculations. The tetragonal phase of Mn$_3$Sn is found to have the lowest total energy among the three phases, similar to Mn$_3$Ga and Mn$_3$Ge. However, the cubic phase of Mn$_3$Sn has higher energy than its hexagonal phase, in contrast to the Mn$_3$Ga and Mn$_3$Ge compounds. In the case of Mn$_3$Sn, we suggest it is difficult for the hexagonal phase to transform from the cubic phase into the tetragonal one. This might explain the lack of the tetragonal $Mn_3Sn$ phase.

**II Calculation details**

First-principles calculations were carried out using Vienna *ab* initio Simulation Package (VASP) [26]. The ions were described using projector augmented wave (PAW) potentials [27]. The generalized gradient approximation (GGA) in the Perdew-Bruke-Ernzerhof (PBE) form [28] was adopted to describe the exchange-correlation interactions between electrons. An energy cutoff of 500 eV was used for the plane wave basis. Spin-orbit coupling was employed in all calculations to describe the non-collinear spin polarization and magnetocrystalline anisotropy. The volume and shape (c/a) of the cubic, tetragonal and hexagonal structures were fully relaxed to get the stable structural configurations with the lowest energy. In addition, the full potential linear augmented plane wave (FLAPW) method [29] was also used to check the validity of the pseudopotential calculations.

**III. Results and discussion**

As shown in Figure 1, the cubic phase belongs to the $X_2YZ$ full Heusler structure (Fm-3m). $Mn_{II}$ (X position), $Mn_I$ (Y position) and Z (Z=Ga, Sn and Ge) atoms occupy the (1/4, 1/4, 1/4), (1/2, 1/2, 1/2) and (0, 0, 0) sites, respectively. The magnetic moments of $Mn_I$ and $Mn_{II}$ orient oppositely. The tetragonal phase (I4/mmm) has an elongated *c* axis and shortened *a* axis as compared to the cubic lattice. The magnetic order, as in the cubic phase, is ferrimagnetic. Our optimized lattice parameters and magnetic moments (per $Mn_3Z$ unit) agree well with previous experiments [1, 3, 21, 24] and calculations [17, 23] (see Tables I and II). The PMA energy is defined as the energy difference between the easy direction (001) and the in-plane direction (010). In the tetragonal phase, the PMA is around 1 meV for all three compounds, consistent with previous calculations on $Mn_3Ga$ and $Mn_3Ge$ [11, 21].

The Mn atoms in the hexagonal phase exhibit various possible magnetic configurations. On the one hand, in the plane of the triangle lattice, the magnetic moments of Mn atoms may point in-plane or out-of-plane. On the other hand, one finds that the Mn-Mn bonds between neighboring layers in of Mn$_3$Z are a little shorter than the in-plane Mn-Mn bonds. Therefore the interlayer magnetic coupling is also important. As a summary, we display the most important configurations in Fig. 2. The AFM type of in-plane ordering in Fig. 2(d) is found to be the most stable. However, the directions of the magnetic moments are not equally separated by 120°, therefore they do not fully cancel each other, resulting in a weak ferromagnetic phase [30~32]. The calculated magnetic moments also agree with previous experiments for all three materials. For example, the magnetic moments are 0.03 $\mu_B$ (exp. 0.045 $\mu_B$ [1]) for Mn$_3$Ga, 0.01 $\mu_B$ (exp. 0.009 $\mu_B$ [2]) for Mn$_3$Sn and 0.01 $\mu_B$ (exp. 0.06 $\mu_B$ [3]) for Mn$_3$Ge. In Mn$_3$Ge, the variation between the experimental and calculated results originates mostly from the fact that the experimental compositions are off-stoichiometry [3, 22].

It is important to compare the energetic stability of the three phases for Mn$_3$Z (Z=Ga, Sn and Ge). Figure 3 shows the total energy dependence on the volume of hexagonal, cubic and tetragonal structures for these three compounds. We relaxed all the structure parameters (volume and shape (c/a) of the lattice) to reach the most stable structures. In the tetragonal and hexagonal phases, for example, the c/a-ratio was fully optimized for any given value of the volume. One unambiguously finds in all three compounds that the tetragonal phase is energetically more stable than the cubic and hexagonal phases. For Mn$_3$Ga and Mn$_3$Ge, the hexagonal phase has the highest total energy, while the cubic phase exhibits the highest energy for Mn$_3$Sn. Although the cubic phase is claimed to be unstable in experimental work [23], its relative energy between these three phases is very important in understanding the stability of the tetragonal and hexagonal phases.

We propose that the hexagonal phase does not change into the tetragonal phase directly; rather it transitions through the cubic lattice, which is structurally intermediate. It is simple to see that distortion along the *c* direction of the cubic phase will lead to the tetragonal phase, while the transition between the cubic and hexagonal phases is not as straightforward. If the cubic lattice is projected along the diagonal direction, one obtains a trianglar lattice. In order to recover a hexagonal phase, four atomic layers including three Mn layers and one Z layer should be compressed into one layer. However, this cannot be realized by a simple projection of ABC sites, for the Z atom overlaps with one Mn atom in this way (Figs. 1(d) and 1(e)). In order to host the additional Mn atoms inside a layer, the honeycomb lattice of Mn (Fig. 1(d)) should change into a Kagome lattice (Figs. 1(f) and 1(g)) to create more available sites for Mn atoms. On the one hand, the transition from the cubic to the hexagonal phase requires compressive pressure along the cubic diagonal direction in order to push Mn and Z atoms inside a plane. On the other hand, the transition from the hexagonal to the cubic phase also needs compressive strain in the *ab* plane of the hexagonal lattice.

For $Mn_3Ga$ and $Mn_3Ge$, the cubic phase is energetically between the hexagonal and tetragonal phases. In this case, the hexagonal phase can easily pass through the cubic phase into the tetragonal phase. Therefore, both phases can be realized in experiments [1, 3, 23, 24]. In contrast to the above two compounds, $Mn_3Sn$ has a cubic phase whose energy is higher than the hexagonal phase. Consequently, it is difficult to transform from an existing hexagonal phase into a tetragonal one, possibly explaining why tetragonal $Mn_3Sn$ has not been synthesized so far.

We can explain the transition between the hexagonal and tetragonal phases using the cubic phase. However, there still remains a puzzle: the tetragonal structure of $Mn_3Sn$ has lower energy than the hexagonal one in our calculations, while experiments only observe the hexagonal phase. This may be due to the structural disorder or off-stoichiometry compositions that commonly exist in experiments [2]. In these cases, the hexagonal phase may have lower energy than the tetragonal counterpart. We will investigate these effects in a future work.

**IV. Conclusions**

We calculated the structures and compared the stabilities of the cubic, hexagonal and tetragonal phases for $Mn_3Ga$, $Mn_3Ge$ and $Mn_3Sn$. The cubic phase plays an important role as an intermediate state in the phase transition from the hexagonal to the tetragonal phase. Consequently, the cubic phase is necessary to analyse the stability of the other two phases and understand the experiments. For $Mn_3Ga$ and $Mn_3Ge$, the cubic phase lies between the hexagonal phase with the highest energy and the tetragonal phase with the lowest energy. Consequently, it is possible to transform the hexagonal phase, through the cubic phase, into the tetragonal phase. This is consistent with experimental observations. However, for Mn$_3$Sn, the cubic phase has a higher energy than the hexagonal phase. Although the tetragonal phase has the lowest energy, it turns out to be difficult to transform the hexagonal phase into the tetragonal one. This agrees with the fact that only hexagonal Mn$_3$Sn has been synthesized in experiments so far. We also propose that external pressure can be utilized to assist the phase transition. In addition, the weak ferromagnetism of hexagonal compounds and the perpendicular magnetocrystalline anisotropy in the tetragonal compounds are consistent with previous experiments.

**Acknowledgements**

We acknowledge the funding support by the ERC Advanced Grant (291472) and the kind help from Dr. J. Karel.

Figure 1 (D. L. Zhang et al.)

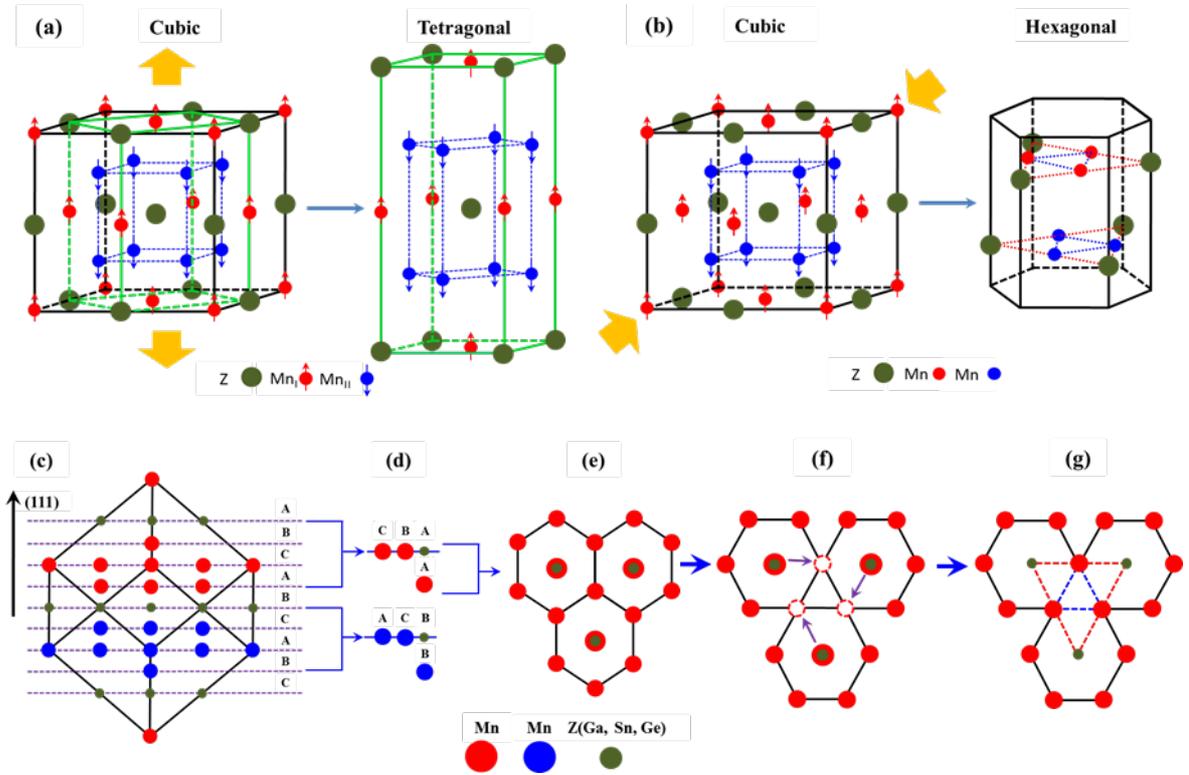

Figure 1 (Colour online) Crystal structure relationships of (a) cubic-to-tetragonal and (b) cubic-to-hexagonal phases. The projection from the cubic to hexagonal structures is shown in (c)-(g). Four atomic layers (three Mn layers and one Z layer) will be projected into a single layer, in order to form the hexagonal structure. The orange arrows in (a)-(b) indicate the direction of structural deformation, and the arrows on the Mn atoms in (a) denote the direction of magnetic moment. In addition, Mn atoms are represented by red or blue balls, and Z (Z=Ga, Sn and Ge) atoms by green balls.

Figure 2 (D. L. Zhang et al.)

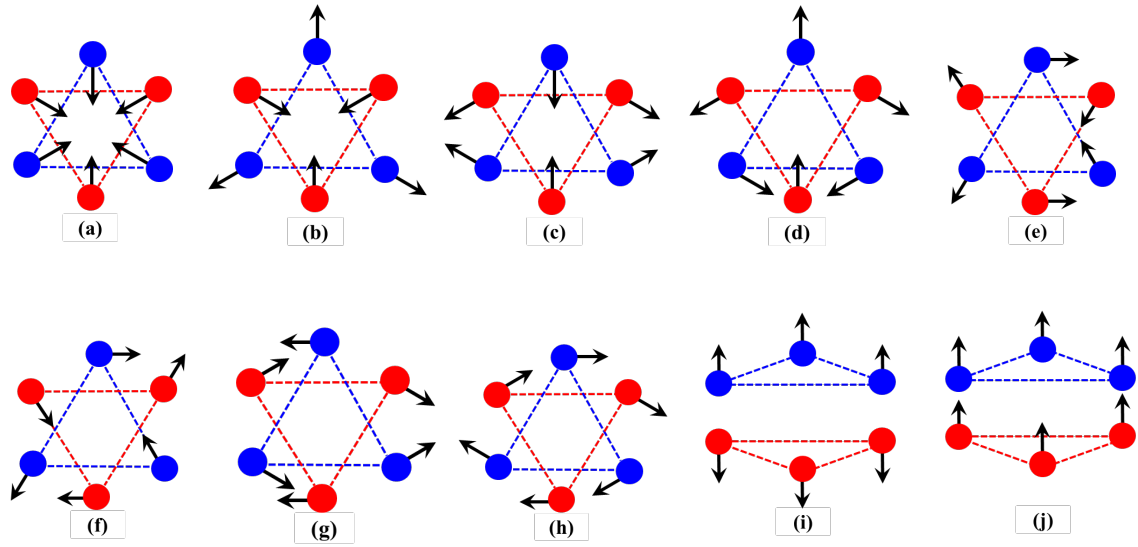

Figure 2 (Colour online) A schematic of possible magnetic configurations in the hexagonal lattice of $Mn_3Z$ (Z=Ga, Sn and Ge). The arrows denote the direction of the magnetic moments. The magnetic moments of Mn atoms lie in the hexagonal basal plane in (a)-(h), and out-of-plane in (i)-(j). The blue and red balls represent the Mn atoms in different planes. The magnetic configuration of (d) is found to be the most stable by our calculations.

Figure 3 (D. L. Zhang et al.)

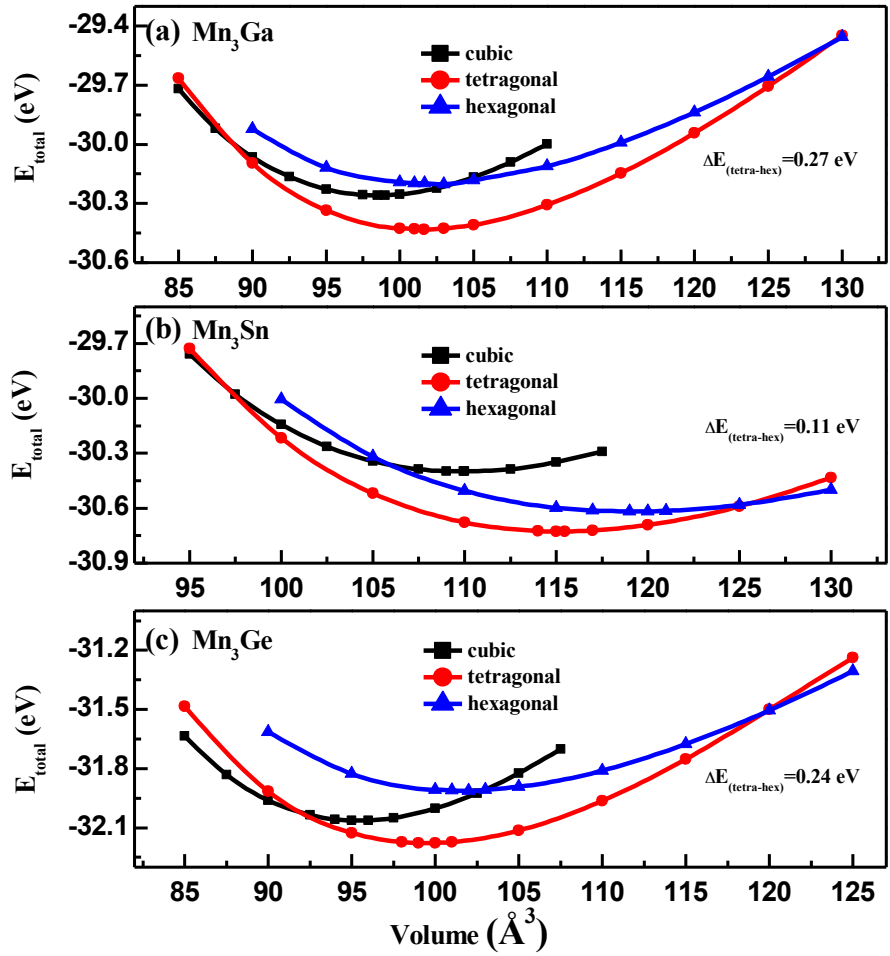

Figure 3 (Colour online) The dependence of the total energy per $Mn_3Z$ unit on volume for the cubic, tetragonal and hexagonal structures of (a) $Mn_3Ga$, (b) $Mn_3Sn$, and (c) $Mn_3Ge$. For a given volume, the shape of the lattice (c/a) is also fully optimized.

Table I. Optimized lattice parameters of hexagonal, cubic, and tetragonal structures in $Mn_3Z$ (Z=Ga, Sn and Ge) Heusler compounds.

|  | Lattice constant (Å) | | |
| --- | --- | --- | --- |
|  | $Mn_3Ga$ | $Mn_3Sn$ | $Mn_3Ge$ |
| Hexagonal | a=5.26, c=4.26 (a=5.400, c=4.353 exp.[1]) | a=5.57, c=4.43 (a=5.665, c=4.531 exp.[2]) | a=5.28, c=4.22 (a=5.360, c=4.320 exp.[3]) |
| Cubic | a=5.82 (a=5.823 calc.[10]) | a=6.04 | a=5.75 |
| Tetragonal | a=3.77, c=7.16 (a=3.77, c=7.16 calc.[23] a=3.909, c=7.098 exp.[24]) | a=3.93, c=7.47 | a=3.75, c=7.12 (a=3.81, c=7.26 exp.[3, 21]) |

Table II. Total magnetic moments per $Mn_3Z$ unit of hexagonal, cubic, and tetragonal structures. For the hexagonal phase, the $\mu_{Mn}$ denotes the magnetic moment of each Mn atom. The energy of perpendicular magnetocrystalline anisotropy (per primitive unit cell, $Mn_6Z_2$) of the tetragonal structure is shown.

|  | Magnetic Moment ($\mu_B$) | | |
| --- | --- | --- | --- |
|  | $Mn_3Ga$ | $Mn_3Sn$ | $Mn_3Ge$ |
| Hexagonal | 0.03 (0.045 exp.[1]) $\mu_{Mn}$=2.5 (2.4±0.2 exp.[1]) | 0.01 (0.009 exp.[2]) $\mu_{Mn}$=2.9 (3.0 exp.[2]) | 0.01 (0.06 exp.[3]) $\mu_{Mn}$=2.5 (2.4±0.2 exp.[3]) |
| Cubic | -0.01(-0.01 calc.[10]) | -1.00 | -1.00 |
| Tetragonal | 1.78 (1.77 calc.[23]) PMA=1.0 meV (1.0 meV calc.[11]) | 1.04 PMA=1.1 meV | 0.97(1.00 exp.[21]) PMA=1.0 meV (0.8 meV calc.[21]) |